\documentclass[submission,creativecommons]{eptcs}

\providecommand{\event}{HCVS 2021} 
\usepackage{breakurl}             
\usepackage{underscore}           
\usepackage{mathtools} 
\usepackage{todonotes}
\usepackage{listings}
\usepackage{makecell}
\usepackage{xspace}
\usepackage{longtable,tablefootnote,tabularx}
\usepackage[title]{appendix}
\usepackage{color, colortbl}
\usepackage{csvsimple}
\usepackage{ifthen}

\title{Competition Report: \CHCC
}
\author{Grigory Fedyukovich \institute{Florida State University, USA} \and
Philipp R\"ummer
\institute{Uppsala University, Sweden}
}
\def\titlerunning{Competition Report: \CHCC}
\def\authorrunning{Grigory Fedyukovich and Philipp R\"ummer}

\newcommand{\LIA}{LIA-nonlin\xspace}
\newcommand{\LIAlin}{LIA-lin\xspace}
\newcommand{\LIAar}{LIA-nonlin-arrays\xspace}
\newcommand{\LIAarshort}{LIA-nl-arrays\xspace}
\newcommand{\LIAlinar}{LIA-lin-arrays\xspace}
\newcommand{\LRATS}{LRA-TS\xspace}
\newcommand{\LRATSpar}{LRA-TS-par\xspace}
\newcommand{\ADT}{ADT-nonlin\xspace}

\newcommand{\CHC}{CHC-COMP\xspace}
\newcommand{\CHCC}{CHC-COMP-21\xspace}
\newcommand{\CHCCold}{CHC-COMP-20\xspace}
\newcommand{\CHCColdold}{CHC-COMP-19\xspace}

\newcommand{\Score}{\textbf{Score}\xspace}
\newcommand{\CPUtime}{\textbf{CPU time}\xspace}
\newcommand{\Walltime}{\textbf{Wall-clock time}\xspace}

\newcommand{\sat}{\textbf{sat}\xspace}
\newcommand{\unsat}{\textbf{unsat}\xspace}
\newcommand{\unknown}{\textbf{unknown}\xspace}

\newcommand{\StarExec}{\href{https://www.starexec.org}{StarExec}\xspace}

\begin{document}
\maketitle

\begin{abstract}
  \CHCC\footnote{\url{https://chc-comp.github.io/}} is the fourth
  competition of solvers for Constrained Horn Clauses. In this year, 7
  solvers participated at the competition, and were evaluated in 7
  separate tracks on problems in linear integer arithmetic, linear
  real arithmetic, arrays, and algebraic data-types. The competition
  was run in March 2021 using the \StarExec computing
  cluster. This report gives an overview of the competition design,
  explains the organisation of the competition, and presents the
  competition results.
\end{abstract}

\section{Introduction}

Constrained Horn Clauses (CHC) have over the last decade emerged as a
uniform framework for reasoning about different aspects of software
safety~\cite{andrey-pldi,BjornerGMR15}. Constrained Horn clauses form
a fragment of first-order logic, modulo various background theories,
in which models can be constructed effectively with the help of
techniques including model checking, abstract interpretation, or
clause transformation. Horn clauses can be used as an intermediate
verification language that elegantly captures various classes of
systems (e.g., sequential code, programs with functions and
procedures, concurrent programs, or reactive systems) and various
verification methodologies (e.g., the use of state invariants,
verification with the help of contracts, Owicki-Gries-style
invariants, or rely-guarantee methods). Horn solvers can be used as
off-the-shelf back-ends in verification tools, and thus enable
construction of verification systems in a modular way.

\CHCC is the fourth competition of solvers for Constrained Horn
Clauses, a competition affiliated with the 8th Workshop on Horn
Clauses for Verification and Synthesis (HCVS) at ETAPS~2021.  The goal
of \CHC is to compare state-of-the-art tools for Horn solving with
respect to performance and effectiveness on realistic, publicly
available benchmarks.  The deadline for submitting solvers to \CHCC
was March~18 2021, resulting in 7 solvers participating, which were
evaluated in the second half of March~2021. The 7 solvers were
evaluated in 7 separate tracks on problems in linear integer
arithmetic, linear real arithmetic, the theory of arrays, and theories
of algebraic data-types. The results of the competition can be found
in Section~\ref{sec:results} of this report, and were presented at the
(virtual) HCVS workshop on March 28 2021.

\subsection{Acknowledgements}

We would like to thank the HCVS chairs, Bishoksan Kafle and Hossein
Hojjat, for hosting \CHC also in this year!

\CHCC heavily built on the infrastructure developed for the
previous instances of \CHC, run by Arie Gurfinkel, Grigory
Fedyukovich, and Philipp R\"ummer, respectively. Contributors to the
competition infrastructure also include Adrien Champion, Dejan
Jovanovic, and Nikolaj Bj\o rner.

Like in the first three competitions, \CHCC was run on
\StarExec \cite{DBLP:conf/cade/StumpST14}. We are extremely grateful
for the computing resources and evaluation environment provided by
\StarExec, and for the fast and competent support by Aaron Stump and
his team whenever problems occurred. \CHCC would not have been
possible without this!

Philipp R\"ummer is
supported by the Swedish Research Council (VR) under grant 2018-04727,
by the Swedish Foundation for Strategic Research (SSF) under the
project WebSec (Ref.\ RIT17-0011), and by the Knut and Alice
Wallenberg Foundation under the project UPDATE.

\section{Brief Overview of the Competition Design}

\subsection{Competition Tracks}

Three new tracks were introduced in \CHCC (namely, \LIAar, \LRATSpar,
\ADT), leading to altogether 7 tracks:
\begin{itemize}
\item \textbf{\LIA}: benchmarks with at least one non-linear clause,
  and linear integer arithmetic as background theory;
\item \textbf{\LIAlin}: benchmarks with only linear clauses, and
  linear integer arithmetic as background theory;
\item \textbf{\LIAar}: benchmarks with at least one non-linear clause, and the
  combined theory of linear integer arithmetic and arrays as
  background theory;
\item \textbf{\LIAlinar}: benchmarks with only linear clauses, and the
  combined theory of linear integer arithmetic and arrays as
  background theory;
\item \textbf{\LRATS}: benchmarks encoding transition systems, with
  linear real arithmetic as background theory. Benchmarks in this
  track have exactly one uninterpreted relation symbol, and exactly
  three linear clauses encoding initial states, transitions, and error
  states;
\item \textbf{\LRATSpar}: same selection of benchmarks as in \LRATS,
  but 2x4 CPU cores were reserved for each task, and the evaluation
  was done with wall-clock time limit; this yields a setting benefiting
  parallel solvers;
\item \textbf{\ADT}: benchmarks with at least one non-linear clause,
  and the algebraic data-types as background theory.
\end{itemize}

\subsection{Computing Nodes}

Two separate queues on \StarExec were used for the competition, one
queue with 15~nodes for the track \LRATSpar, and one with
20~nodes for all other tracks.  Each node had two quadcore CPUs. In
\LRATSpar, each job was run on its own node during the
competition runs, while in the other tracks each node was used to run
two jobs in parallel.  The
\href{https://www.starexec.org/starexec/public/machine-specs.txt}{machine
  specifications} are:
\begin{verbatim}
Intel(R) Xeon(R) CPU E5-2609 0 @ 2.40GHz (2393 MHZ)
    10240  KB Cache
    263932744 kB main memory

# Software:
OS:       CentOS Linux release 7.7.1908 (Core)
kernel:   3.10.0-1062.4.3.el7.x86_64
glibc:    glibc-2.17-292.el7.x86_64
          gcc-4.8.5-39.el7.x86_64
          glibc-2.17-292.el7.i686
\end{verbatim}

\subsection{Test and Competition Runs}

The solvers submitted to \CHCC were evaluated twice:
\begin{itemize}
\item in a first set of \textbf{test runs}, in which (optional)
  pre-submissions of the solvers were evaluated to check their
  configurations and identify possible inconsistencies. For the test
  runs a smaller set of randomly selected benchmarks was used. 
  In the test runs, each solver-benchmark
  pair was limited to 600s CPU time, 600s wall-clock time, and 64GB
  memory.
\item in the \textbf{competition runs}, the results of which
  determined the outcome of \CHCC. The selection of the benchmarks for
  the competition runs is described in Section~\ref{sec:selection},
  and the evaluation of the competition runs in
  Section~\ref{sec:eval}. In the competition run of \LRATSpar, each
  job was limited to 1800s wall-clock time, and 64GB memory.  In the
  competition runs of all other tracks, each job was limited to 1800s
  CPU time, 1800s wall-clock time, and 64GB memory.
\end{itemize}

\subsection{Evaluation of the Competition Runs}
\label{sec:eval}

The evaluation of the competition runs was in this year done using the
\texttt{summarize.py} script available in the repository
\url{https://github.com/chc-comp/scripts}, and on the basis of the
data provided by \StarExec through the ``job information'' data export
function. The ranking of solvers in each track was done
based on the \Score reached by the solvers in the competition run for
that track.  In case two solvers had equal \Score, the ranking of the
two solvers was determined by \CPUtime (for \LRATSpar, by
\Walltime). It was assumed that the outcome of running one solver on
one benchmark can only be \sat, \unsat, or \unknown; the last outcome
includes solvers giving up, running out of resources, or crashing.

\medskip
The definition of \Score, \CPUtime, and \Walltime are:
\begin{itemize}
\item \Score: the number of \sat or \unsat results produced by a
  solver on the benchmarks of a track.
\item \CPUtime: the total CPU time needed by a solver to produce its
  answers in some track, including \unknown answers.
\item \Walltime: the total wall-clock time needed by a solver to
  produce its answers in some track, including \unknown answers.
\end{itemize}

In addition, the following feature is included in the results for each
solver and each track:
\begin{itemize}
\item \textbf{\#unique}: The number of \sat or \unsat results produced
  by a solver for benchmarks for which all other solvers returned
  \unknown.
\end{itemize}

We decided to not include the \textbf{Space} feature, specifying the
total maximum virtual memory consumption, in the tables, since this
number is less telling for solvers running in a JVM.

\section{Competition Benchmarks}

\subsection{File Format}

\CHC represents benchmarks in a fragment of the SMT-LIB 2.6
format. The fragment is defined on
\url{https://chc-comp.github.io/format.html}.  The conformance of a
well-typed SMT-LIB script with the \CHC fragment can be checked using
the \texttt{format-checker} available on
\url{https://github.com/chc-comp/chc-tools}.

\subsection{Benchmark Processing in Tracks other than \ADT}
\label{sec:processing}

All benchmarks used in \CHCC were pre-processed using the
\texttt{format.py} script available in the repository
\url{https://github.com/chc-comp/scripts}, using the command line
\begin{verbatim}
> python3 format.py --out_dir <outdir> --merge_queries True <smt-file>
\end{verbatim}
The script tries to translate arbitrary Horn-like problems in SMT-LIB
format to problems within the \CHC fragment.  Only benchmarks
processed in this way were used in the competition.

The option \verb!--merge_queries! has the effect of merging multiple
queries in a benchmark into a single query by introducing an auxiliary
nullary predicate. This transformation was introduced in \CHCCold, and
is discussed in \cite{DBLP:journals/corr/abs-2008-02939}.

After processing with \texttt{format.py}, benchmarks were checked and
categorised into the four tracks using the \texttt{format-checker}
scripts available on
\url{https://github.com/chc-comp/chc-tools}.

Benchmarks that could not be processed by \texttt{format.py} were
rejected by the \texttt{format-checker}.
Benchmarks that did not conform to any of
the competition tracks, were not used in \CHCC.

\subsection{Benchmark Processing in \ADT}
\label{sec:processingADT}

Benchmarks used in the \ADT track were preprocessed by eliminating all theory constraints and recursively-defined functions. The transformation was performed using the feature of the \textsc{RInGen} tool~\cite{kostyukov2021finite}.
This way, we were able to satisfy the input-language constraints for all four tools entering the competition in this track.
In the future, we, however, plan introducing other ADT-related tracks with benchmarks over ADT and linear arithmetic and/or arrays.

\subsection{Benchmark Inventory}

\begin{table}[tb]
  \begin{center}\footnotesize
  \newcommand{\empt}{\multicolumn{2}{c}{}}
  \begin{tabular}{l*{6}{r@{ / }l}}
    \textbf{Repository} &
               \multicolumn{2}{c}{\LIA} &
               \multicolumn{2}{c}{\LIAlin} &
               \multicolumn{2}{c}{\LIAar} &
               \multicolumn{2}{c}{\LIAlinar} &
               \multicolumn{2}{c}{\LRATS} &
               \multicolumn{2}{c}{\ADT}
    \\\hline\hline
    adt-purified & \empt & \empt &  \empt & \empt & \empt & 67 & 67
    \\\hline
    aeval & \empt & 54&54
    \\\hline
    eldarica-misc & 69&66 & 147&134 
    \\\hline
    extra-small-lia & \empt & 55&55 
    \\\hline
    hcai & 135&133 & 100&86 & 25&25 &  39&39 
    \\\hline
    hopv & 68&67 & 49&48 
    \\\hline
    jayhorn &  5138&5084 & 75&73 
    \\\hline
    kind2 & 851&738 
    \\\hline
    ldv-ant-med & \empt & \empt & 79&79 & 10&10
    \\\hline
    ldv-arrays & \empt & \empt & 821&546 &3&2
    \\\hline
    llreve & 59&57 & 44&44 &\empt& 31&31 
    \\\hline
    quic3 & \empt & \empt &\empt& 43&43 
    \\\hline
    ringen & \empt & \empt & \empt & \empt & \empt & 439 & 439
    \\\hline
    sally & \empt & \empt & \empt &\empt& 177 & 174
    \\\hline
    seahorn & 68&66 & 3396&2822 
    \\\hline
    synth/nay-horn & 119&114
    \\\hline
    synth/semgus & \empt & \empt & 5371*&4839*
    \\\hline
    tricera & 4&4 & 405&405 
    \\\hline
    vmt & \empt & 905&802 & \empt &\empt& 99 & 98
    \\\hline\hline
    chc-comp19 & 271&265 & 325&313 & 15&15 & 290&290 & 228&226
    \\\hline
    sv-comp & 1643&1169 & 3150&2932 & 855&779 & 79&73 
    \\\hline\hline
    \textbf{Total} & ~~8425&7763~~ & ~~8705&7768~~ & ~~7166 &6283 & ~~495&488 & 504&498 & ~~~506 & 506
  \end{tabular}
  \end{center}

  \caption{Summary of benchmarks available on
    \url{https://github.com/chc-comp} and in the
    \href{https://www.starexec.org/starexec/secure/explore/spaces.jsp?id=73700}{StarExec
      CHC space}.  For each collection of benchmarks and each \CHCC
    track, the first number gives the total number of benchmarks, and
    the second number the number of contributed unique benchmarks
    (after discarding duplicate benchmarks). In the benchmark family
    synth/semgus, only 2357/2331 benchmarks were taken into account
    for the competition, as processing of the other benchmarks
    (according to Section~\ref{sec:processing}) was incorrect due to
    inconsistent filename conventions. This mistake was only
    discovered after the main competition runs were finished, and
    could not be corrected in time.}
  \label{tab:benchmarks}
\end{table}

In contrast to most other competitions, \CHC stores benchmarks in a
decentralised way, in multiple repositories managed by the
contributors of the benchmarks themselves.  Table~\ref{tab:benchmarks}
summarises the number of benchmarks that were obtained by collecting
benchmarks from all available repositories using the process in
Section~\ref{sec:processing} and
Section~\ref{sec:processingADT}. Duplicate benchmarks were identified
by computing a checksum for each (processed) benchmark, and were
discarded.

The repository chc-comp19-benchmarks of benchmarks selected for
\CHCColdold was included in the collection, because this repository
contains several unique families of benchmarks that are not available
in other repositories under \url{https://github.com/chc-comp}. Such
benchmarks include problems generated by the Ultimate tools in the
\LIAlinar track.

From jayhorn-benchmarks, only the problems generated for sv-comp-2020
were considered, which subsume the problems for sv-comp-2019.

For \ADT, benchmarks originate from the TIP suite (originally, designed for theorem-proving) and verification of programs in functional languages.

\section{Benchmark Rating and Selection}
\label{sec:selection}

\begin{table}[tb]
  \begin{center}\footnotesize
  \newcommand{\empt}{\multicolumn{3}{c}{}}
  \begin{tabular}{l*{3}{r@{ / }r@{ / }r}}
    &          \multicolumn{3}{c}{~~~~~~~\LIA} &
               \multicolumn{3}{c}{~~~~~~~~\LIAlin} &
               \multicolumn{3}{c}{~~~~~\LIAarshort}
                                          \\
    \textbf{Repository} &
               \#A & \#B & \#C &
               \#A & \#B & \#C &
               ~~~~\#A & \#B & \#C
    \\\hline\hline
    aeval & \empt & 11&15&28
    \\\hline
    eldarica-misc & 35&4&27 & 105&20&9
    \\\hline
    extra-small-lia & \empt & 21&24&10
    \\\hline
    hcai & 74&44&15 & 73&8&5 & 14&6&5
    \\\hline
    hopv & 60&7& & 47&1&
    \\\hline
    jayhorn &  2688&769&1627 & 73&
    \\\hline
    kind2 & 250&455&33
    \\\hline
    ldv-ant-med & \empt & \empt & &25&54
    \\\hline
    ldv-arrays & \empt & \empt & &127&419
    \\\hline
    llreve & 35&13&9 & 37&5&2
    \\\hline
    seahorn & 38&19&9 & 977&985&860
    \\\hline
    synth/nay-horn & 46&30&38
    \\\hline
    synth/semgus & \empt & \empt & 282&768&1281
    \\\hline
    tricera & 4&& & 28&14&363
    \\\hline
    vmt & \empt & 85&616&101
    \\\hline\hline
    chc-comp19 & 144&80&41 & 80&101&132 &&7&8
    \\\hline
    sv-comp & 1013&144&12 & 2801&17&114 & 258&268&253
    \\\hline\hline
    \textbf{Total} & ~~~~4387&1565&1811 & ~~~~~4338&1806&1624 & 554&1201&2020
  \end{tabular}
  \end{center}

  \caption{The number of unique benchmarks with ratings
    A / B / C, respectively.}
  \label{tab:rating}
\end{table}

\begin{table}[tb]
  \begin{center}\footnotesize
  \newcommand{\empt}{\multicolumn{2}{c}{}}
  \begin{tabular}{l*{3}{@{~~~~~~~~~~~~}r@{~~~~}r}@{\qquad}*{3}{r}}
    &          \multicolumn{2}{l}{\hspace*{-2ex}\LIA} &
               \multicolumn{2}{l}{\LIAlin} &
               \multicolumn{2}{l}{\hspace*{-3ex}\LIAarshort} &
               ~~\LIAlinar & \LRATS & \ADT
                                          \\
    \textbf{Repository} &
               $N_r$ & \#Sel & 
               $N_r$ & \#Sel & 
               $N_r$ & \#Sel & \#Selected~~~~ & \#Selected & \#Selected~~
    \\\hline\hline
    adt-purified & \empt & \empt & \empt &&& 67~~~~~~~~
    \\\hline
    aeval & \empt & 10&30
    \\\hline
    eldarica-misc & 10&30 & 15&39
    \\\hline
    extra-small-lia & \empt & 10&30
    \\\hline
    hcai & 20&55 & 15&28 & 5&15 & 39~~~~~~~~
    \\\hline
    hopv & 10&17 & 10&11
    \\\hline
    jayhorn &  30&90 & 10&10
    \\\hline
    kind2 & 30&90
    \\\hline
    ldv-ant-med & \empt & \empt & 20&60  & 10~~~~~~~~
    \\\hline
    ldv-arrays & \empt & \empt & 30&90 &2~~~~~~~~
    \\\hline
    llreve & 15&37 & 10&17 & \empt & 31~~~~~~~~
    \\\hline
    quic3 & \empt & \empt & \empt & 43~~~~~~~~
    \\\hline
    ringen & \empt & \empt & \empt & & & 111~~~~~~~~
    \\\hline
    sally & \empt & \empt & \empt & & 174~~~~~
    \\\hline
    seahorn & 15&39 & 30&90
    \\\hline
    synth/nay-horn & 20&60
    \\\hline
    synth/semgus & 20&60 & \empt & 45&135
    \\\hline
    tricera & 1&1 & 20&60
    \\\hline
    vmt & \empt & 30&90 && \empt & 98~~~~~
    \\\hline\hline
    chc-comp19 & 30&90 & 30&90 & 5&15 & 290~~~~~~~~ & 226~~~~~
    \\\hline
    sv-comp & 30&72 & 30&90 & 45&135 & 73~~~~~~~~
    \\\hline\hline
    \textbf{Total} & &581 & &585 & & 450 & 488~~~~~~~~ & 498~~~~~ & 178~~~~~~~~
  \end{tabular}
  \end{center}

  \caption{The number of selected unique benchmarks for the \CHCC
    tracks.}
  \label{tab:selectionNum}
\end{table}

This section describes how the benchmarks for \CHCC were selected
among the unique benchmarks summarised in
Table~\ref{tab:benchmarks}. For the competition tracks \LIAlinar, 
\LRATS, and \ADT, the benchmark library only contains 488, 498, and 506 unique
benchmarks, respectively, which are small enough sets to use all
benchmarks in the competition. For the tracks \LIA, \LIAlin, and \LIAar, in
contrast, too many benchmarks are available, so that a representative
sample of the benchmarks had to be chosen. 

To gauge the difficulty of the available problems in \LIA, \LIAlin,
\LIAar, a simple rating based on the performance of the \CHCCold
solvers was computed. The same approach was used in the last
competition, \CHCCold, using solvers from \CHCColdold. In this year,
the two top-ranked competing solvers from \CHCCold were run for a few
seconds on each of the benchmarks:\footnote{Run on an Intel
  Core~i5-650 2-core machine with 3.2GHz. All timeouts are in terms of
  wall-clock time.}
\begin{itemize}
\item For \textbf{\LIA} and \textbf{\LIAlin}: Spacer (timeout 5s) and
  Eldarica-abs (timeout 10s);
\item For \textbf{\LIAar}: Spacer (timeout 5s) and Ultimate Unihorn
  (timeout 10s). Since \LIAar was not evaluated at \CHCCold, the
  top-ranked solvers from the track \LIAlinar were chosen.
\end{itemize}
All solvers were run using the same binary and same options as in
\CHCCold. For the JVM-based tools, Eldarica-abs and Ultimate Unihorn,
the higher timeout was chosen to compensate for the JVM start-up delay.

The outcomes of those test runs gave rise to three possible ratings
for each benchmark:
\begin{itemize}
\item \textbf{A:} both tools were able to determine the benchmark
  status within the given time budget.
\item \textbf{B:} only one tool could determine the benchmark status.
\item \textbf{C:} both tools timed out.
\end{itemize}

The number of benchmarks per rating are shown in
Table~\ref{tab:rating}. As can be seen from the table, the simple
rating method separates the benchmarks into partitions of comparable
size, and provides some information about the relative hardness of the
problems in the different repositories.

From each repository~$r$, up to $3\cdot N_r$ benchmarks were then
selected randomly: $N_r$ benchmarks with rating~A, $N_r$ benchmarks
with rating~B, and $N_r$ benchmarks with rating~C. If a repository
contained fewer than $N_r$ benchmarks for some particular rating,
instead benchmarks with the next-higher rating were chosen. As special
cases, up to $N_r$ benchmarks were selected from repositories with
only A-rated benchmarks; up to $2\cdot N_r$ benchmarks from
repositories with only B-rated benchmarks; and up to $3\cdot N_r$
benchmarks from repositories with only C-rated benchmarks.

The number~$N_r$ was chosen individually for each repository, based on
a manual inspection of the repository to judge the diversity of the
contained benchmarks. The chosen~$N_r$, and the numbers of selected
benchmarks for each repository, are given in
Table~\ref{tab:selectionNum}.

For the actual selection of benchmarks with rating~X, the following
Unix command was used:
\begin{verbatim}
> cat <rating-X-benchmark-list> | sort -R | head -n <num>
\end{verbatim}

\medskip
The final set of benchmarks selected for \CHCC can be found in the
github repository
\url{https://github.com/chc-comp/chc-comp21-benchmarks}, and on
\StarExec in the public space \linebreak
\href{https://www.starexec.org/starexec/secure/explore/spaces.jsp?id=442514}{\texttt{CHC/CHC-COMP/chc-comp21-benchmarks}}.

\section{Solvers Entering \CHCC}

\begin{table}[tb]
  \begin{center}
    \small
  \begin{tabularx}{\linewidth}{X*{7}{X}}
    \hline
    \textbf{Solver} &  \LIA & \LIAlin & \LIAar & \LIAlinar & \LRATS  & \LRATSpar  & \ADT
    \\\hline\hline
    Golem & --- &  \ttfamily LIA-Lin & --- & --- & \ttfamily LRA-TS & \ttfamily LRA-TS & ---
    \\\hline
    PCSat & \ttfamily pcsat\_tb\_ucore\_ar &  \ttfamily pcsat\_tb\_ucore\_ar & --- & --- & --- & --- & \ttfamily pcsat\_tb\_ucore\_reduce\_quals
    \\\hline
    Spacer
           & \ttfamily LIA\-NONLIN
           & \ttfamily LIA-LIN
           &\ttfamily   LIA\-NONLIN\-ARRAYS
           &\ttfamily  LIA-LIN\-ARRAYS
           & \ttfamily LRA-TS
           & \ttfamily LRA-TS
           & \ttfamily ADT\-NONLIN
    \\\hline
    \raggedright
    Ultimate TreeAutomizer & \ttfamily default & \ttfamily default & \ttfamily default & \ttfamily default & \ttfamily default & \ttfamily default & ---
    \\\hline
    \raggedright
    Ultimate Unihorn & \ttfamily default & \ttfamily default & \ttfamily default & \ttfamily default & \ttfamily default & \ttfamily default & ---
    \\\hline
    RInGen & --- & ---&---&---&---&---& \ttfamily default
    \\\hline\hline
    \raggedright
    Eldarica (Hors Concours) & \ttfamily def   & \ttfamily def & \ttfamily def & \ttfamily def & --- & --- & \ttfamily def
    \\\hline
  \end{tabularx}
  \end{center}
  \caption{The submitted solvers, and the configurations used in the
  individual tracks.}
  \label{tab:solvers}
\end{table}

In total, 7~solvers were submitted to \CHCC: 6 competing solvers, and
one further solver (Eldarica, co-developed by one of the competition
organisers) that was entering outside of the competition.  A summary of
the participating solvers is given in Table~\ref{tab:solvers}.

More details about the participating solvers are provided in the
solver descriptions in Section~\ref{sec:solvers}. The binaries of the
solvers used for the competition runs can be found in the public
\StarExec space \linebreak
\href{https://www.starexec.org/starexec/secure/explore/spaces.jsp?id=442514}{\texttt{CHC/CHC-COMP/chc-comp21-benchmarks}}.

\section{Competition Results}
\label{sec:results}

The winners and top-ranked solvers of the seven \CHCC tracks are:
\begin{center}
  \small
  \begin{tabularx}{\linewidth-2ex}{l@{}|*{7}{X}}
    \hline
    & \raisebox{-1.5ex}{\LIA} & \raisebox{-1.5ex}{\LIAlin} & \LIAarshort & \LIAlinar & \raisebox{-1.5ex}{\LRATS} & \LRATSpar & \ADT
    \\\hline\hline
    \textbf{Winner}~~~ & \textbf{Spacer} & \textbf{Spacer} & \textbf{Spacer} & \textbf{Spacer} & \textbf{Spacer} & \textbf{Spacer} & \textbf{Spacer}
    \\\hline
    Place 2& Ultimate Unihorn & Golem & \raggedright Ultimate Unihorn & Ultimate Unihorn & Golem & Golem & RInGen
    \\\hline
    Place 3 & \raggedright PCSat & \raggedright Ultimate Unihorn & Ultimate TreeAutomizer & Ultimate TreeAutomizer & Ultimate TreeAutomizer & Ultimate TreeAutomizer & PCSat
    \\\hline
  \end{tabularx}
\end{center}

Detailed results for the seven tracks are provided in the tables on
page~\pageref{tab:results-LIA}.

\subsection{Observed Issues and Fixes during the Competition Runs}

\paragraph{Fixes in Spacer:}

During the competition runs, it was observed that Spacer, in the
version submitted by March~18, did not run correctly on \StarExec and
did not produce output for any of the benchmarks. Since this issue was
discovered soon after the start of the competition runs, the
organisers decided to let the Spacer authors submit a corrected
version. The problem turned out to be compilation/linking-related, and
the results presented in this report were produced with the fixed
version of Spacer.  To ensure fairness of the competition, all teams
were given time until March~20 to submit revised versions of their
tools.

\paragraph{Fixes in Golem:}

One case of inconsistent results was observed in the competition runs
in the track \LIAlin. For the benchmark \verb!chc-LIA-Lin_502.smt2!,
the tool Golem reported \unsat, while Spacer and Eldarica reported
\sat. The author of Golem confirmed that the inconsistency was due to
a bug in the loop acceleration in Golem, and could provide a corrected
version in which loop acceleration was switched off. The results
presented in this report were produced with this fixed version of
Golem.

To ensure fairness of the competition, we provide the following table
comparing the results of the two versions of Golem in track
\LIAlin. The table shows that the fix in Golem led to marginally worse
performance of the solver, and therefore did not put the other
solvers at an unfair disadvantage:
\begin{center}
  \begin{tabular}{l*{2}{@{\qquad}c}}
    \hline
    & \#sat & \#unsat
    \\\hline\hline
    Golem (original) & 185 & 133
    \\\hline
    Golem (fixed) &  179 & 133
    \\\hline
  \end{tabular}
\end{center}

\section{Conclusions}
\label{sec:next}

The organisers would like to congratulate the general winner of this
year's \CHC, the solver \textbf{Spacer}, as well as all solvers and
tool authors for their excellent performance!  Thanks go to everybody
who has been helping with infrastructure, scripts, benchmarks, or in
other ways, see the acknowledgements in the introduction; and to the
HCVS workshop for hosting \CHC!

\bigskip
\noindent
The organisers also identified several questions and issues that
should be discussed and addressed in the next editions, in order to
keep \CHC an interesting and relevant competition:
\begin{itemize}
\item \textbf{Models and counterexamples} (as already discussed in
  \cite{DBLP:journals/corr/abs-2008-02939}). A concern brought up
  again at the HCVS workshop is the generation of models and/or
  counterexample certificates, highlighting the user demand for this
  functionality. Since at the moment many tools do not support
  certificates yet, this could initially happen in the scope of a new
  track, or by awarding a higher number of points for each produced
  and verified model/counterexample.
\item \textbf{Multi-query benchmarks} (as already discussed in
  \cite{DBLP:journals/corr/abs-2008-02939}). We propose to extend the
  \CHC fragment of SMT-LIB to include also problems with multiple
  queries. This would leave the decision how to handle multi-query
  benchmarks to each solver. For solvers that can only solve problems
  with a single query, a script is available to transform multi-query
  problems to single-query problems.
\item \textbf{The \LRATS track.} This restricted track was created to
  enable also solvers that only support traditional transition systems
  to enter. However, no such solver was submitted to \CHCC (in
  contrast to \CHCCold), which means that the results presented in
  this report do not fully reflect the state of the art for such
  problems. For future instances of \CHC, it can be considered to
  replace \LRATS with a general LRA track, dropping the restriction to
  problems in transition system form.
\item\textbf{\ADT}: As mentioned in Sect.~\ref{sec:processingADT}, the syntactic restrictions on the ADT tasks were needed to let more solvers participate in the competition. Thus, we had only ``pure ADT'' problems. However, as the technology evolves, we expect more solvers to participate in the next editions of the competition. Thus,  ADT tasks that also use constraints in other theories (if collected in sufficient amounts) could form new tracks.
\item \textbf{A bigger set of benchmarks is needed, and all users and
    tool authors are encouraged to submit benchmarks!} In particular,
  in the \LIA, \LRATS, and \ADT tracks, the competition results indicate that
  more and/or harder benchmarks are required.
\end{itemize}

\newpage
\section{Solver Descriptions}
\label{sec:solvers}

The tool descriptions in this section were contributed by the tool
submitters, and the copyright on the texts remains with the individual
authors.

\makeatletter
\renewcommand\paragraph{\@startsection{paragraph}{4}{\z@}%
                                    {2ex \@plus1ex \@minus.2ex}%
                                    {-1em}%
                                    {\normalfont\normalsize\bfseries}}
\makeatother

\newcommand{\toolname}[1]{\bigskip\section*{#1}}
\newcommand{\toolsubmitter}[2]{%
  \noindent
  \begin{tabular}[t]{@{}l@{}}
    #1\\#2
  \end{tabular}\hspace*{8ex}}
\newcommand{\toolalgorithm}{\paragraph{Algorithm.}}
\newcommand{\toolnew}{\paragraph{News in 2021.}}
\newcommand{\toolarchitecture}{\paragraph{Architecture and Implementation.}}
\newcommand{\toolconfiguration}{\paragraph{Configuration in \CHCC.}}
\newcommand{\toollink}[2]{\par\medskip\noindent\url{#1}\\#2}





\newcommand{\golem}{\textsc{Golem}}
\toolname{\golem}


\toolsubmitter{Martin Blicha}{Universit\`{a} della Svizzera italiana, Switzerland}

\toolalgorithm
\golem{} is a new CHC solver, still under active development.
It can solve systems of linear clauses with Linear Real or Integer Arithmetic as the background theory and it is able to provide witnesses for both satisfiable and unsatisfiable systems.

Its current reasoning engine is a re-implementation of the \textsc{Impact} algorithm~\cite{McMillan2006} and thus falls into the category of interpolation-based model-checking approaches.

\toolarchitecture

\golem{} is implemented in C++ and built on top of the interpolating SMT solver \textsc{OpenSMT}~\cite{OpenSMT2} which is used for both satisfiability solving and interpolation. The only dependencies are those inherited from  \textsc{OpenSMT}: Flex, Bison and GMP libraries.

\toolconfiguration
\golem{} was run with its default settings, except that its experimental loop acceleration module had to be disabled, because it contained a bug in the submitted version. Note that the SMT theory needs to be specified.

\texttt{\$ golem --logic QF\_LRA --accelerate-loops=false}

\texttt{\$ golem --logic QF\_LIA --accelerate-loops=false}

\toollink{http://verify.inf.usi.ch/golem}{MIT LICENSE}

\toolname{PCSat}

\toolsubmitter{Yu Gu}{University of Tsukuba, Japan}
\toolsubmitter{Hiroshi Unno}{University of Tsukuba, Japan}

\toolalgorithm

PCSat is a solver for a general class of second-order constraints.  Its applications include but not limited to branching-time temporal verification, relational verification, dependent refinement type inference, program synthesis, and infinite-state game solving.

PCSat is based on CounterExample-Guided Inductive Synthesis (CEGIS), with the support of multiple synthesis engines including template-based~\cite{Unno2021}, decision-tree-based~\cite{Kura2021}, and graphical-model-based~\cite{Satake2020} ones.

\toolarchitecture

PCSat is designed and implemented as a highly-configurable solver, allowing us to test various combinations of synthesis engines, example sampling methods, template refinement strategies, and qualifier generators.  This design is enabled by a powerful module system and metaprogramming features of the OCaml functional programming language.  PCSat uses Z3 as the backend SMT solver.

\toolnew

We supported the theory of algebraic datatypes and implemented a preprocessor for eliminating irrelevant arguments of predicates.

\toolconfiguration

PCSat is run with the solver configuration file ``pcsat_tb_ucore_ar.json'' in the LIA-Nonlin and LIA-Lin tracks and ``pcsat_tb_ucore_reduce_quals.json'' in the ADT-Nonlin track.  Both configurations enable the template-based synthesis engine.

\toollink{https://github.com/hiroshi-unno/coar}{Apache License 2.0}


\newcommand{\ourtool}{\textsc{RInGen}}
\newcommand{\cvc}{\textsc{CVC4}}

\toolname{\ourtool{} v1.1}


\toolsubmitter{Yurii Kostyukov}{Saint Petersburg State University, JetBrains Research, Russia}\\ 
\toolsubmitter{Dmitry Mordvinov}{Saint Petersburg State University, JetBrains Research, Russia}

\toolalgorithm
\ourtool{} stands for a \emph{R}egular \emph{In}variant \emph{Gen}erator, where
 \emph{regular invariants}~\cite{kostyukov2021finite} are represented by \emph{finite tree automata}.
While invariant representations based on first-order logic (FOL) can only access finitely many subterms, regular invariants have an ability to ``scan'' an ADT term to the unbounded depth via automaton rules.
Tree automata also enjoy useful decidability properties and the corresponding regular tree languages are closed under all set operations, which makes regular invariants a promising alternative to FOL-based invariant representations.

\ourtool{} rewrites a system of CHCs over ADTs into a formula over uninterpreted function symbols by eliminating all disequalities, testers, and selectors from the clause bodies. Then the satisfiability modulo theory of ADTs is reduced to satisfiability modulo theory of uninterpreted functions with equality (EUF). After that, an off-the-shelf finite model finder is applied to build a finite model of the reduced verification conditions. Finally, using the correspondence between finite models and tree automata, the automaton representing the safe inductive invariant of the original system is obtained.
Full algorithmic details of the \ourtool{} can be found in~\cite{kostyukov2021finite}.

\toolarchitecture

\ourtool{} accepts input in the SMTLIB2 format and produces CHCs over pure ADT sorts in SMTLIB2 and Prolog.
It takes conditions with a property as input and checks if the property holds, returning SAT and a safe inductive invariant, or terminates with UNSAT otherwise.
We exploit \cvc{} (using \texttt{cvc4 -{}-finite-model-find}) at the backend to find regular models.
Besides regular models, a finite model finding approach of \cvc{}~\cite{reynolds2013finite} v1.8 based on quantifier instantiation provides us with sound satisfiability checking.



\toolconfiguration

The tool is built and run with the following arguments:\\
\centerline{\texttt{solve --timelimit \$tlimit --quiet --output-directory "\$dir" cvc4f "\$input"}.}

\toollink{https://github.com/Columpio/RInGen/releases/tag/v1.1}{BSD 3-Clause License}




\newcommand{\Spacer}{\textsc{Spacer}\xspace}
\newcommand{\GSpacer}{\textsc{GSpacer}\xspace}
\newcommand{\pob}{\textsc{pob}s\xspace}
\newcommand{\pobs}{\textsc{pob}s\xspace}
\newcommand{\false}{\emph{false}\xspace}

\toolname{\Spacer}


\toolsubmitter{Hari Govind V K}{University of Waterloo, Canada}
\toolsubmitter{Arie Gurfinkel}{University of Waterloo, Canada}

\toolalgorithm

\Spacer~\cite{DBLP:journals/fmsd/KomuravelliGC16} is an IC3/PDR-style algorithm
for solving linear and non-linear CHCs. Given a set of CHCs, it iteratively
proves the unreachability of \false at larger and larger depths until a model is
found or the set of CHCs is proven unsatisfiable. To prove unreachability at a
particular depth, \Spacer recursively generates sets of predecessor
states~(called proof obligations~(\pobs)) from which \false can be derived and
blocks them. Once a \pob is blocked, \Spacer generalizes the proof to learn a
\emph{lemma} that blocks multiple \pobs. \Spacer uses many heuristics to learn
lemmas. These include interpolation, inductive generalization and quantifier
generalization. The latest version of Spacer presents a new heuristic for
learning lemmas~\cite{gspc, DBLP:conf/iccad/KFG20}.

The current implementation of \Spacer supports linear and non-linear CHCs in the
theory of Arrays, Linear Arithmetic, FixedSizeBitVectors, and Algebraic Data Types. \Spacer can
generate both quantified and quantifier free models as well as resolution proof
of unsatisfiability.

\toolarchitecture
\Spacer is implemented on top of the \textsc{Z3} theorem prover. It uses many SMT solvers implemented in \textsc{Z3}. Additionally, it implements an interpolating SMT solver.

\toolconfiguration


\Spacer has several configurations. The following options are common to all configurations:
\begin{verbatim}
        fp.xform.tail_simplifier_pve=false fp.validate=true 
        fp.spacer.mbqi=false fp.spacer.use_iuc=true 
\end{verbatim}
To activate global guidance~\cite{gspc}, we use the following options:
\begin{verbatim}
        fp.spacer.global=true fp.spacer.concretize=true 
        fp.spacer.conjecture=true fp.spacer.expand_bnd=true
\end{verbatim}
To activate quantifier generalization~\cite{DBLP:conf/atva/GurfinkelSV18}, we use:
\begin{verbatim}
        fp.spacer.q3.use_qgen=true fp.spacer.q3.instantiate=true 
        fp.spacer.q3=true fp.spacer.ground_pobs=false
\end{verbatim}
In the arithmetic tracks~(LRA-TS, LIA-LIN, LIA-NONLIN), we
ran two threads in parallel. The first thread ran \Spacer with
global guidance. The second thread ran \textsc{Z3}'s BMC engine:
\begin{verbatim}
        fp.engine=bmc
\end{verbatim}
In the array tracks~(LIA-LIN-ARRAYS, LIA-NONLIN-ARRAYS), we again ran two
threads in parallel. The first thread had both global guidance and quantifier generalization. The second thread had only quantifier generalization.
In the ADT tracks~(ADT-LIN, ADT-NONLIN), we ran one thread which used only
global guidance. Additionally, for the ADT tracks, we turned off one of the optimizations in \Spacer:
\begin{verbatim}
        fp.spacer.use_inc_clause=false
\end{verbatim}

\toollink{https://github.com/Z3Prover/z3}{MIT License}

\toolname{Ultimate TreeAutomizer 0.1.25-6b0a1c7}


\toolsubmitter{Matthias Heizmann}{University of Freiburg, Germany}
\toolsubmitter{Daniel Dietsch}{University of Freiburg, Germany}

\medskip
\toolsubmitter{Jochen Hoenicke}{University of Freiburg, Germany}
\toolsubmitter{Alexander Nutz}{University of Freiburg, Germany}

\medskip
\toolsubmitter{Andreas Podelski}{University of Freiburg, Germany}

\toolalgorithm

The \textsc{Ultimate TreeAutomizer} solver implements an approach that is based on tree automata~\cite{journals/corr/abs-1907-03998}.
In this approach potential counterexamples to satisfiability are considered as a regular set of trees.
In an iterative \nobreak{CEGAR} loop we analyze potential counterexamples.
Real counterexamples lead to an \textit{unsat} result.
Spurious counterexamples are generalized to a regular set of spurious counterexamples
and subtracted from the set of potential counterexamples that have to be considered.
In case we detected that all potential counterexamples are spurious, the result is \textit{sat}.
The generalization above is based on tree interpolation and
regular sets of trees are represented as tree automata.


\toolarchitecture

\textsc{TreeAutomizer} is a toolchain in the 
\textsc{Ultimate} framework.
This toolchain first parses the CHC input and then runs the \texttt{treeautomizer} plugin which
implements the above mentioned algorithm.
We obtain tree interpolants from the SMT solver SMTInterpol%
\footnote{\url{https://ultimate.informatik.uni-freiburg.de/smtinterpol/}}%
~\cite{cade/HoenickeS18}.
For checking satisfiability, we use the
 Z3 SMT solver%
\footnote{\url{https://github.com/Z3Prover/z3}}%
.
The tree automata are implemented in \textsc{Ultimate}'s automata library%
\footnote{\url{https://ultimate.informatik.uni-freiburg.de/automata_library}}%
.
The \textsc{Ultimate} framework is written in Java and build upon the Eclipse Rich Client Platform (RCP). The source code is available at
GitHub\footnote{\url{https://github.com/ultimate-pa/}}.


\toolconfiguration

Our StarExec archive for the competition is shipped with the \texttt{bin/starexec\_run\_default}
shell script calls the \textsc{Ultimate} command line interface with the\linebreak
\texttt{TreeAutomizer.xml} toolchain file and
the \texttt{TreeAutomizerHopcroftMinimization.epf} settings file.
Both files can be found in toolchain (resp. settings) folder of \textsc{Ultimate}'s repository.


\toollink{https://ultimate.informatik.uni-freiburg.de/}{LGPLv3 with a linking exception for Eclipse RCP}

\newpage

\toolname{Ultimate Unihorn 0.1.25-6b0a1c7}


\toolsubmitter{Matthias Heizmann}{University of Freiburg, Germany}
\toolsubmitter{Daniel Dietsch}{University of Freiburg, Germany}

\medskip
\toolsubmitter{Jochen Hoenicke}{University of Freiburg, Germany}
\toolsubmitter{Alexander Nutz}{University of Freiburg, Germany}

\medskip
\toolsubmitter{Andreas Podelski}{University of Freiburg, Germany}

\toolalgorithm

\textsc{Ultimate Unihorn} reduces the satisfiability problem for a set of constraint Horn clauses
to a software verfication problem.
In a first step \textsc{Unihorn} applies a 
yet unpublished translation in which the constraint Horn clauses
are translated into a recursive program
that is nondeterministic and
whose correctness is specified by an assert statement
The program is correct (i.e., no execution violates the assert statement)
if and only if the set of CHCs is satisfiable.
For checking whether the recursive program satisfies its specification,
Unihorn uses \textsc{Ultimate Automizer}~\cite{tacas/HeizmannCDGHLNM18}
which implements an automata-based approach to software verification~\cite{cav/HeizmannHP13}.


\toolarchitecture

\textsc{Ultimate Unihorn} is a toolchain in the 
\textsc{Ultimate} framework.
This toolchain first parses the CHC input and then runs the \texttt{chctoboogie} plugin which
does the translation from CHCs into a recursive program.
We use the Boogie
language to represent that program.
Afterwards the default toolchain for verifying a recursive Boogie programs by \textsc{Ultimate Automizer} is applied.
The \textsc{Ultimate} framework shares the libraries for handling SMT formulas with the SMTInterpol SMT solver.
While verifying a program, \textsc{Ultimate Automizer} needs SMT solvers
for checking satisfiability,
for computing Craig interpolants and
for computing unsatisfiable cores.
The version of \textsc{Unihorn} that participated in the competition
used the SMT solvers SMTInterpol%
\footnote{\url{https://ultimate.informatik.uni-freiburg.de/smtinterpol/}}%
and Z3%
\footnote{\url{https://github.com/Z3Prover/z3}}%
.
The \textsc{Ultimate} framework is written in Java and build upon the Eclipse Rich Client Platform (RCP). The source code is available at
GitHub\footnote{\url{https://github.com/ultimate-pa/}}.


\toolconfiguration

Our StarExec archive for the competition is shipped with the \texttt{bin/starexec\_run\_default}
shell script calls the \textsc{Ultimate} command line interface with the\linebreak
\texttt{AutomizerCHC.xml} toolchain file and
the \texttt{AutomizerCHC\_No\_Goto.epf} settings file.
Both files can be found in toolchain (resp. settings) folder of \textsc{Ultimate}'s repository.


\toollink{https://ultimate.informatik.uni-freiburg.de/}{LGPLv3 with a linking exception for Eclipse RCP}

\newpage

\toolname{Eldarica v2.0.6 (Hors Concours)}


\toolsubmitter{Zafer Esen}{Uppsala University, Sweden}
\toolsubmitter{Hossein Hojjat}{University of Tehran, Iran}
\toolsubmitter{Philipp R\"ummer}{Uppsala University, Sweden}

\toolalgorithm

Eldarica~\cite{FMCAD2018HojjatRummer} is a Horn solver applying
classical algorithms from model checking: predicate abstraction and
counterexample-guided abstraction refinement (CEGAR).  Eldarica can
solve Horn clauses over linear integer arithmetic, arrays, algebraic
data-types, and bit-vectors.  It can process Horn clauses and programs
in a variety of formats, implements sophisticated algorithms to solve
tricky systems of clauses without diverging, and offers an elegant API
for programmatic use.

\toolarchitecture

Eldarica is entirely implemented in Scala, and only depends on Java or
Scala libraries, which implies that Eldarica can be used on any
platform with a JVM. For computing abstractions of systems of Horn
clauses and inferring new predicates, Eldarica invokes the SMT solver
Princess~\cite{princess08} as a library.

\toolnew

Compared to the last competition, Eldarica now uses a new array solver in
the tracks \LIAar and \LIAlinar.

\toolconfiguration

Eldarica is in the competition run with the option \verb!-abstractPO!,
which enables a simple portfolio mode: two instances of the solver are
run in parallel, one with the default options, and one with the option
\verb!-abstract:off! to switch off the interpolation abstraction
technique.

\toollink{https://github.com/uuverifiers/eldarica}{BSD licence}


\newpage
\bibliographystyle{eptcs}
\bibliography{refs}

\newcommand{\noop}[1]{}
\begin{thebibliography}{10}
\providecommand{\bibitemdeclare}[2]{}
\providecommand{\surnamestart}{}
\providecommand{\surnameend}{}
\providecommand{\urlprefix}{Available at }
\providecommand{\url}[1]{\texttt{#1}}
\providecommand{\href}[2]{\texttt{#2}}
\providecommand{\urlalt}[2]{\href{#1}{#2}}
\providecommand{\doi}[1]{doi:\urlalt{http://dx.doi.org/#1}{#1}}
\providecommand{\bibinfo}[2]{#2}

\bibitemdeclare{inproceedings}{BjornerGMR15}
\bibitem{BjornerGMR15}
\bibinfo{author}{Nikolaj \surnamestart Bj{\o}rner\surnameend},
  \bibinfo{author}{Arie \surnamestart Gurfinkel\surnameend},
  \bibinfo{author}{Kenneth~L. \surnamestart McMillan\surnameend} \&
  \bibinfo{author}{Andrey \surnamestart Rybalchenko\surnameend}
  (\bibinfo{year}{2015}): \emph{\bibinfo{title}{{Horn} Clause Solvers for
  Program Verification}}.
\newblock In: {\sl \bibinfo{booktitle}{Fields of Logic and Computation {II} -
  Essays Dedicated to Yuri Gurevich on the Occasion of His 75th Birthday}}, pp.
  \bibinfo{pages}{24--51}, \doi{10.1007/978-3-319-23534-9_2}.

\bibitemdeclare{inproceedings}{journals/corr/abs-1907-03998}
\bibitem{journals/corr/abs-1907-03998}
\bibinfo{author}{Daniel \surnamestart Dietsch\surnameend},
  \bibinfo{author}{Matthias \surnamestart Heizmann\surnameend},
  \bibinfo{author}{Jochen \surnamestart Hoenicke\surnameend},
  \bibinfo{author}{Alexander \surnamestart Nutz\surnameend} \&
  \bibinfo{author}{Andreas \surnamestart Podelski\surnameend}
  (\bibinfo{year}{2019}): \emph{\bibinfo{title}{Ultimate TreeAutomizer
  {(CHC-COMP} Tool Description)}}.
\newblock In: {\sl \bibinfo{booktitle}{HCVS/PERR@ETAPS}}, {\sl
  \bibinfo{series}{{EPTCS}}} \bibinfo{volume}{296}, pp.
  \bibinfo{pages}{42--47}, \doi{10.4204/EPTCS.296.7}.

\bibitemdeclare{inproceedings}{andrey-pldi}
\bibitem{andrey-pldi}
\bibinfo{author}{Sergey \surnamestart Grebenshchikov\surnameend},
  \bibinfo{author}{Nuno~P. \surnamestart Lopes\surnameend},
  \bibinfo{author}{Corneliu \surnamestart Popeea\surnameend} \&
  \bibinfo{author}{Andrey \surnamestart Rybalchenko\surnameend}
  (\bibinfo{year}{2012}): \emph{\bibinfo{title}{Synthesizing Software Verifiers
  from Proof Rules}}.
\newblock In: {\sl \bibinfo{booktitle}{PLDI}}, \bibinfo{publisher}{ACM}, pp.
  \bibinfo{pages}{405--416}, \doi{10.1145/2254064.2254112}.

\bibitemdeclare{inproceedings}{DBLP:conf/atva/GurfinkelSV18}
\bibitem{DBLP:conf/atva/GurfinkelSV18}
\bibinfo{author}{Arie \surnamestart Gurfinkel\surnameend},
  \bibinfo{author}{Sharon \surnamestart Shoham\surnameend} \&
  \bibinfo{author}{Yakir \surnamestart Vizel\surnameend}
  (\bibinfo{year}{2018}): \emph{\bibinfo{title}{Quantifiers on Demand}}.
\newblock In \bibinfo{editor}{Shuvendu~K. \surnamestart Lahiri\surnameend} \&
  \bibinfo{editor}{Chao \surnamestart Wang\surnameend}, editors: {\sl
  \bibinfo{booktitle}{Automated Technology for Verification and Analysis - 16th
  International Symposium, {ATVA} 2018, Los Angeles, CA, USA, October 7-10,
  2018, Proceedings}}, {\sl \bibinfo{series}{Lecture Notes in Computer
  Science}} \bibinfo{volume}{11138}, \bibinfo{publisher}{Springer}, pp.
  \bibinfo{pages}{248--266}, \doi{10.1007/978-3-030-01090-4\_15}.

\bibitemdeclare{inproceedings}{tacas/HeizmannCDGHLNM18}
\bibitem{tacas/HeizmannCDGHLNM18}
\bibinfo{author}{Matthias \surnamestart Heizmann\surnameend},
  \bibinfo{author}{Yu{-}Fang \surnamestart Chen\surnameend},
  \bibinfo{author}{Daniel \surnamestart Dietsch\surnameend},
  \bibinfo{author}{Marius \surnamestart Greitschus\surnameend},
  \bibinfo{author}{Jochen \surnamestart Hoenicke\surnameend},
  \bibinfo{author}{Yong \surnamestart Li\surnameend},
  \bibinfo{author}{Alexander \surnamestart Nutz\surnameend},
  \bibinfo{author}{Betim \surnamestart Musa\surnameend},
  \bibinfo{author}{Christian \surnamestart Schilling\surnameend},
  \bibinfo{author}{Tanja \surnamestart Schindler\surnameend} \&
  \bibinfo{author}{Andreas \surnamestart Podelski\surnameend}
  (\bibinfo{year}{2018}): \emph{\bibinfo{title}{Ultimate Automizer and the
  Search for Perfect Interpolants - (Competition Contribution)}}.
\newblock In: {\sl \bibinfo{booktitle}{{TACAS} {(2)}}}, {\sl
  \bibinfo{series}{LNCS}} \bibinfo{volume}{10806},
  \bibinfo{publisher}{Springer}, pp. \bibinfo{pages}{447--451},
  \doi{10.1007/978-3-319-89963-3_30}.

\bibitemdeclare{inproceedings}{cav/HeizmannHP13}
\bibitem{cav/HeizmannHP13}
\bibinfo{author}{Matthias \surnamestart Heizmann\surnameend},
  \bibinfo{author}{Jochen \surnamestart Hoenicke\surnameend} \&
  \bibinfo{author}{Andreas \surnamestart Podelski\surnameend}
  (\bibinfo{year}{2013}): \emph{\bibinfo{title}{Software Model Checking for
  People Who Love Automata}}.
\newblock In: {\sl \bibinfo{booktitle}{{CAV}}}, {\sl \bibinfo{series}{LNCS}}
  \bibinfo{volume}{8044}, \bibinfo{publisher}{Springer}, pp.
  \bibinfo{pages}{36--52}, \doi{10.1007/978-3-642-39799-8_2}.

\bibitemdeclare{inproceedings}{cade/HoenickeS18}
\bibitem{cade/HoenickeS18}
\bibinfo{author}{Jochen \surnamestart Hoenicke\surnameend} \&
  \bibinfo{author}{Tanja \surnamestart Schindler\surnameend}
  (\bibinfo{year}{2018}): \emph{\bibinfo{title}{Efficient Interpolation for the
  Theory of Arrays}}.
\newblock In: {\sl \bibinfo{booktitle}{{IJCAR}}}, {\sl \bibinfo{series}{LNCS}}
  \bibinfo{volume}{10900}, \bibinfo{publisher}{Springer}, pp.
  \bibinfo{pages}{549--565}, \doi{10.1007/978-3-319-94205-6_36}.

\bibitemdeclare{inproceedings}{FMCAD2018HojjatRummer}
\bibitem{FMCAD2018HojjatRummer}
\bibinfo{author}{Hossein \surnamestart Hojjat\surnameend} \&
  \bibinfo{author}{Philipp \surnamestart R{\"{u}}mmer\surnameend}
  (\bibinfo{year}{2018}): \emph{\bibinfo{title}{The {ELDARICA} Horn Solver}}.
\newblock In \bibinfo{editor}{Nikolaj \surnamestart Bj{\o}rner\surnameend} \&
  \bibinfo{editor}{Arie \surnamestart Gurfinkel\surnameend}, editors: {\sl
  \bibinfo{booktitle}{2018 Formal Methods in Computer Aided Design, {FMCAD}}},
  \bibinfo{publisher}{{IEEE}}, pp. \bibinfo{pages}{1--7},
  \doi{10.23919/FMCAD.2018.8603013}.

\bibitemdeclare{inproceedings}{OpenSMT2}
\bibitem{OpenSMT2}
\bibinfo{author}{Antti E.~J. \surnamestart Hyv{\"a}rinen\surnameend},
  \bibinfo{author}{Matteo \surnamestart Marescotti\surnameend},
  \bibinfo{author}{Leonardo \surnamestart Alt\surnameend} \&
  \bibinfo{author}{Natasha \surnamestart Sharygina\surnameend}
  (\bibinfo{year}{2016}): \emph{\bibinfo{title}{OpenSMT2: An {SMT} Solver for
  Multi-core and Cloud Computing}}.
\newblock In: {\sl \bibinfo{booktitle}{Theory and Applications of
  Satisfiability Testing - {SAT} 2016 - 19th International Conference,
  Bordeaux, France, July 5-8, 2016, Proceedings}}, {\sl
  \bibinfo{series}{Lecture Notes in Computer Science}} \bibinfo{volume}{9710},
  \bibinfo{publisher}{Springer}, pp. \bibinfo{pages}{547--553},
  \doi{10.1007/978-3-319-40970-2\_35}.

\bibitemdeclare{inproceedings}{gspc}
\bibitem{gspc}
\bibinfo{author}{Hari Govind~V \surnamestart K\surnameend},
  \bibinfo{author}{YuTing \surnamestart Chen\surnameend},
  \bibinfo{author}{Sharon \surnamestart Shoham\surnameend} \&
  \bibinfo{author}{Arie \surnamestart Gurfinkel\surnameend}
  (\bibinfo{year}{2020}): \emph{\bibinfo{title}{Global Guidance for Local
  Generalization in Model Checking}}.
\newblock In: {\sl \bibinfo{booktitle}{Computer Aided Verification - 32nd
  International Conference, {CAV} 2020, Los Angeles, CA, USA, July 21-24, 2020,
  Proceedings, Part {II}}}, {\sl \bibinfo{series}{Lecture Notes in Computer
  Science}} \bibinfo{volume}{12225}, \bibinfo{publisher}{Springer}, pp.
  \bibinfo{pages}{101--125}, \doi{10.1007/978-3-030-53291-8\_7}.

\bibitemdeclare{inproceedings}{DBLP:conf/iccad/KFG20}
\bibitem{DBLP:conf/iccad/KFG20}
\bibinfo{author}{Hari Govind~V. \surnamestart K.\surnameend},
  \bibinfo{author}{Grigory \surnamestart Fedyukovich\surnameend} \&
  \bibinfo{author}{Arie \surnamestart Gurfinkel\surnameend}
  (\bibinfo{year}{2020}): \emph{\bibinfo{title}{Word Level Property Directed
  Reachability}}.
\newblock In: {\sl \bibinfo{booktitle}{{IEEE/ACM} International Conference On
  Computer Aided Design, {ICCAD} 2020, San Diego, CA, USA, November 2-5,
  2020}}, \bibinfo{publisher}{{IEEE}}, pp. \bibinfo{pages}{107:1--107:9},
  \doi{10.1145/3400302.3415708}.

\bibitemdeclare{article}{DBLP:journals/fmsd/KomuravelliGC16}
\bibitem{DBLP:journals/fmsd/KomuravelliGC16}
\bibinfo{author}{Anvesh \surnamestart Komuravelli\surnameend},
  \bibinfo{author}{Arie \surnamestart Gurfinkel\surnameend} \&
  \bibinfo{author}{Sagar \surnamestart Chaki\surnameend}
  (\bibinfo{year}{2016}): \emph{\bibinfo{title}{{SMT-based Model Checking for
  Recursive Programs}}}.
\newblock {\sl \bibinfo{journal}{Formal Methods Syst. Des.}}
  \bibinfo{volume}{48}(\bibinfo{number}{3}), pp. \bibinfo{pages}{175--205},
  \doi{10.1007/s10703-016-0249-4}.

\bibitemdeclare{inproceedings}{kostyukov2021finite}
\bibitem{kostyukov2021finite}
\bibinfo{author}{Yurii \surnamestart Kostyukov\surnameend},
  \bibinfo{author}{Dmitry \surnamestart Mordvinov\surnameend} \&
  \bibinfo{author}{Grigory \surnamestart Fedyukovich\surnameend}
  (\bibinfo{year}{2021}): \emph{\bibinfo{title}{Beyond the Elementary
  Representations of Program Invariants over Algebraic Data Types}}.
\newblock In: {\sl \bibinfo{booktitle}{{PLDI} '21: 42nd {ACM} {SIGPLAN}
  International Conference on Programming Language Design and Implementation,
  Virtual Event, Canada, June 20-25, 20211}}, \bibinfo{publisher}{{ACM}}, pp.
  \bibinfo{pages}{451--465}, \doi{10.1145/3453483.3454055}.

\bibitemdeclare{inproceedings}{Kura2021}
\bibitem{Kura2021}
\bibinfo{author}{Satoshi \surnamestart Kura\surnameend},
  \bibinfo{author}{Hiroshi \surnamestart Unno\surnameend} \&
  \bibinfo{author}{Ichiro \surnamestart Hasuo\surnameend}
  (\bibinfo{year}{2021}): \emph{\bibinfo{title}{Decision Tree Learning in
  CEGIS-Based Termination Analysis}}.
\newblock In: {\sl \bibinfo{booktitle}{Computer Aided Verification - 33rd
  International Conference, {CAV} 2021, Virtual Event, July 20-23, 2021,
  Proceedings, Part {II}}}, {\sl \bibinfo{series}{Lecture Notes in Computer
  Science}} \bibinfo{volume}{12760}, \bibinfo{publisher}{Springer}, pp.
  \bibinfo{pages}{75--98}, \doi{10.1007/978-3-030-81688-9\_4}.

\bibitemdeclare{inproceedings}{McMillan2006}
\bibitem{McMillan2006}
\bibinfo{author}{Kenneth~L. \surnamestart McMillan\surnameend}
  (\bibinfo{year}{2006}): \emph{\bibinfo{title}{Lazy Abstraction with
  Interpolants}}.
\newblock In \bibinfo{editor}{Thomas \surnamestart Ball\surnameend} \&
  \bibinfo{editor}{Robert~B. \surnamestart Jones\surnameend}, editors: {\sl
  \bibinfo{booktitle}{Computer Aided Verification, 18th International
  Conference, {CAV} 2006, Seattle, WA, USA, August 17-20, 2006, Proceedings}},
  {\sl \bibinfo{series}{Lecture Notes in Computer Science}}
  \bibinfo{volume}{4144}, \bibinfo{publisher}{Springer}, pp.
  \bibinfo{pages}{123--136}, \doi{10.1007/11817963\_14}.

\bibitemdeclare{inproceedings}{reynolds2013finite}
\bibitem{reynolds2013finite}
\bibinfo{author}{Andrew \surnamestart Reynolds\surnameend},
  \bibinfo{author}{Cesare \surnamestart Tinelli\surnameend},
  \bibinfo{author}{Amit \surnamestart Goel\surnameend} \& \bibinfo{author}{Sava
  \surnamestart Krsti{\'c}\surnameend} (\bibinfo{year}{2013}):
  \emph{\bibinfo{title}{{Finite model finding in SMT}}}.
\newblock In: {\sl \bibinfo{booktitle}{International Conference on Computer
  Aided Verification}}, {\sl \bibinfo{series}{Lecture Notes in Computer
  Science}} \bibinfo{volume}{8044}, \bibinfo{publisher}{Springer}, pp.
  \bibinfo{pages}{640--655}, \doi{10.1007/978-3-642-39799-8\_42}.

\bibitemdeclare{inproceedings}{princess08}
\bibitem{princess08}
\bibinfo{author}{Philipp \surnamestart R{\"u}mmer\surnameend}
  (\bibinfo{year}{2008}): \emph{\bibinfo{title}{A Constraint Sequent Calculus
  for First-Order Logic with Linear Integer Arithmetic}}.
\newblock In: {\sl \bibinfo{booktitle}{Proceedings, 15th International
  Conference on Logic for Programming, Artificial Intelligence and Reasoning}},
  {\sl \bibinfo{series}{LNCS}} \bibinfo{volume}{5330},
  \bibinfo{publisher}{Springer}, pp. \bibinfo{pages}{274--289},
  \doi{10.1007/978-3-540-89439-1_20}.

\bibitemdeclare{inproceedings}{DBLP:journals/corr/abs-2008-02939}
\bibitem{DBLP:journals/corr/abs-2008-02939}
\bibinfo{author}{Philipp \surnamestart R{\"{u}}mmer\surnameend}
  (\bibinfo{year}{2020}): \emph{\bibinfo{title}{Competition Report:
  {CHC-COMP-20}}}.
\newblock In \bibinfo{editor}{Laurent \surnamestart Fribourg\surnameend} \&
  \bibinfo{editor}{Matthias \surnamestart Heizmann\surnameend}, editors: {\sl
  \bibinfo{booktitle}{Proceedings 8th International Workshop on Verification
  and Program Transformation and 7th Workshop on Horn Clauses for Verification
  and Synthesis, VPT/HCVS@ETAPS 2020, Dublin, Ireland, 25-26th April 2020}},
  {\sl \bibinfo{series}{{EPTCS}}} \bibinfo{volume}{320}, pp.
  \bibinfo{pages}{197--219}, \doi{10.4204/EPTCS.320.15}.

\bibitemdeclare{inproceedings}{Satake2020}
\bibitem{Satake2020}
\bibinfo{author}{Yuki \surnamestart Satake\surnameend},
  \bibinfo{author}{Hiroshi \surnamestart Unno\surnameend} \&
  \bibinfo{author}{Hinata \surnamestart Yanagi\surnameend}
  (\bibinfo{year}{2020}): \emph{\bibinfo{title}{Probabilistic Inference for
  Predicate Constraint Satisfaction}}.
\newblock In: {\sl \bibinfo{booktitle}{The Thirty-Fourth {AAAI} Conference on
  Artificial Intelligence, {AAAI} 2020, The Thirty-Second Innovative
  Applications of Artificial Intelligence Conference, {IAAI} 2020, The Tenth
  {AAAI} Symposium on Educational Advances in Artificial Intelligence, {EAAI}
  2020, New York, NY, USA, February 7-12, 2020}}, \bibinfo{publisher}{{AAAI}
  Press}, pp. \bibinfo{pages}{1644--1651}, \doi{10.1609/aaai.v34i02.5526}.

\bibitemdeclare{inproceedings}{DBLP:conf/cade/StumpST14}
\bibitem{DBLP:conf/cade/StumpST14}
\bibinfo{author}{Aaron \surnamestart Stump\surnameend}, \bibinfo{author}{Geoff
  \surnamestart Sutcliffe\surnameend} \& \bibinfo{author}{Cesare \surnamestart
  Tinelli\surnameend} (\bibinfo{year}{2014}): \emph{\bibinfo{title}{StarExec:
  {A} Cross-Community Infrastructure for Logic Solving}}.
\newblock In \bibinfo{editor}{St{\'{e}}phane \surnamestart Demri\surnameend},
  \bibinfo{editor}{Deepak \surnamestart Kapur\surnameend} \&
  \bibinfo{editor}{Christoph \surnamestart Weidenbach\surnameend}, editors:
  {\sl \bibinfo{booktitle}{Automated Reasoning - 7th International Joint
  Conference, {IJCAR}}}, {\sl \bibinfo{series}{LNCS}} \bibinfo{volume}{8562},
  \bibinfo{publisher}{Springer}, pp. \bibinfo{pages}{367--373},
  \doi{10.1007/978-3-319-08587-6\_28}.

\bibitemdeclare{inproceedings}{Unno2021}
\bibitem{Unno2021}
\bibinfo{author}{Hiroshi \surnamestart Unno\surnameend},
  \bibinfo{author}{Tachio \surnamestart Terauchi\surnameend} \&
  \bibinfo{author}{Eric \surnamestart Koskinen\surnameend}
  (\bibinfo{year}{2021}): \emph{\bibinfo{title}{Constraint-based Relational
  Verification}}.
\newblock In: {\sl \bibinfo{booktitle}{Computer Aided Verification - 33rd
  International Conference, {CAV} 2021, Virtual Event, July 20-23, 2021,
  Proceedings, Part {I}}}, {\sl \bibinfo{series}{Lecture Notes in Computer
  Science}} \bibinfo{volume}{12759}, \bibinfo{publisher}{Springer}, pp.
  \bibinfo{pages}{742--766}, \doi{10.1007/978-3-030-81685-8\_35}.

\end{thebibliography}

\newcommand{\solver}{}
\newcommand{\eldgray}{\ifthenelse{\equal{\solver}{Eldarica}}{\cellcolor{lightgray}}{}}
\newcommand{\readcsv}[1]{%
    \begin{tabular}{>{\eldgray}l*{6}{>{\eldgray}r}}
      \hline
      \bfseries Solver & \bfseries Score & \bfseries \#sat & \bfseries \#unsat &
         \bfseries CPU time/s &
               \bfseries Wall-clock/s & \bfseries \#unique
      \\\hline
      \csvreader[head to column names,late after line=\\\hline]{#1}{}%
      {\solver & \ok & \sat & \uns &  \time & \real & \uniq}
    \end{tabular}
}

\begin{table}[p]
  \caption{Solver performance on the 581 benchmarks of the \LIA\ track}
  \label{tab:results-LIA}

  \smallskip
  { \small\centering
    \readcsv{csv/LIA-NonLin.csv}

  }

  \caption{Solver performance on the 585 benchmarks of the \LIAlin\ track}
  \label{tab:results-LIAlin}

  \smallskip
  { \small\centering
    \readcsv{csv/LIA-Lin.csv}

  }

  \caption{Solver performance on the 450 benchmarks of the \LIAar\ track}
  \label{tab:results-LIAar}

  \smallskip
  { \small\centering
    \readcsv{csv/LIA-NonLin-Arrays.csv}

  }

  \caption{Solver performance on the 488 benchmarks of the \LIAlinar\ track}
  \label{tab:results-LIAlinar}

  \smallskip
  { \small\centering
    \readcsv{csv/LIA-Lin-Arrays.csv}

  }

  \caption{Solver performance on the 498 benchmarks of the \LRATS\ track}
  \label{tab:results-LRATS}

  \smallskip
  { \small\centering
    \readcsv{csv/LRA-TS.csv}

  }

  \caption{Solver performance on the 498 benchmarks of the \LRATSpar\ track}
  \label{tab:results-LRATSpar}

  \smallskip
  { \small\centering
    \readcsv{csv/LRA-TS-par.csv}

  }

  \caption{Solver performance on the 178 benchmarks of the \ADT\ track}
  \label{tab:results-ADT}

  \smallskip
  { \small\centering
    \readcsv{csv/ADT-NonLin.csv}

  }

\end{table}

\end{document}